\documentclass[12pt]{article}
\usepackage[dvips]{epsfig}
\usepackage{graphicx}
\usepackage{latexsym,amsmath,amsfonts,amsthm,amssymb}

\theoremstyle{plain}
\newtheorem{theorem}{Theorem}[section]

\newtheorem{corollary}[theorem]{Corollary}
\newtheorem{lemma}[theorem]{Lemma}

\newcommand{\zbar}{\bar{z}}
\newcommand{\ZZ}{\mathbb{Z}}

\newcommand{\fz}{{\rm{z}}}

\newcommand{\cL}{\mathcal{L}}

\newcommand{\CC}{\mathbb{C}}
\newcommand{\RR}{\mathbb{R}}

\newcommand{\one}{1\!\!1}
\newcommand{\rarrow}{\rightarrow}

\title{Branched Polymers and Dimensional Reduction}

\author{\begin{tabular}{ccc}
    David C. Brydges           & John Z. Imbrie \\
        Department of Mathematics & Department of Mathematics \\
        The University of British Columbia &University of Virginia\\
        Room 121, 1984 Mathematics Road& Charlottesville VA 22904-4137\\
        Vancouver, B.C., Canada V6T 1Z2& {\tt ji2k@virginia.edu}\\
        {\tt db5d@math.ubc.ca}  & \\
        and&\\
Department of Mathematics&\\
       University of Virginia&\\
       Charlottesville VA 22904-4137&\\
\end{tabular}}
       
\date{October 1, 2001}

\begin{document}
\maketitle
\begin{abstract}
We establish an exact
relation between self-avoiding branched polymers in $D+2$ continuum
dimensions and the hard-core continuum gas at negative activity in $D$ dimensions. We
review conjectures and results on critical exponents for $D+2 = 2,3,4$
and show that they are corollaries of our result. We explain the connection (first
proposed by Parisi and Sourlas) between branched polymers in $D+2$ dimensions and
the Yang-Lee edge singularity in $D$ dimensions.
\end{abstract}

\section{Introduction}\label{section:introduction}

A branched polymer is usually defined \cite{Sla99} to be a finite
subset $\{y_{1},\dotsc ,y_{N} \}$ of the lattice $\ZZ ^{D+2}$ together
with a tree graph whose vertices are $\{y_{1},\dotsc ,y_{N} \}$ and
whose edges $\{y_{i},y_{j} \}$ are such that $|y_{i}-y_{j}|=1$ so that
points in an edge of the tree graph are necessarily nearest neighbors.
A tree graph is a connected graph without loops. Since the points
$y_{i}$ are distinct, branched polymers are \emph{self-avoiding}.  The
picture shows a branched polymer with $N=9$ vertices on a
two-dimensional lattice.

\begin{center}
\begin{picture}(203,155)
\linethickness{2pt}

\put(-3.4,-4.75){
 \multiput(0,0)(0,30){6}{
  \multiput(0,15)(30,0){8}{${\cdot}$}
  }
}

\put(120,75){\line(1,0){60}}
\put(120,45){\line(0,1){60}}
\put(90,105){\line(1,0){60}}
\put(90,105){\line(0,1){30}}
\put(150,105){\line(0,1){30}}

\end{picture}
\end{center}

Critical exponents may be defined by considering statistical ensembles
of branched polymers.  Define two branched polymers to be equivalent
when one is a lattice translate of the other, and let $c_{N}$ be the
number of equivalence classes of branched polymers with $N$
vertices. Some authors prefer to consider the number of branched
polymers that contain the origin.  This is $N c_{N}$, since there are $N$ representatives of each
class which contain the origin.  

One expects that $c_{N}$ has an asymptotic law of the form
\[
c_{N} \sim  N^{ -\theta } {\fz}_c^{-N} .
\]
%where $a_{n}\sim b_{n}$ means that $a_{n}/b_{n}$ has a non-zero limit
%as $n\rightarrow \infty$. 
in the sense that $\lim_{N\rightarrow \infty} -\frac{1}{\ln N} \ln [c_N z_c^N] = \theta$.
The \emph{critical exponent} $ \theta$ is
conjectured to be \emph{universal}, meaning that (unlike ${\fz}_c$) it
should be independent of the local structure of the lattice. For
example, it should be the same on a triangular lattice, or in the
continuum model to be considered in this paper.

In 1981 Parisi and Sourlas \cite{PS81} conjectured exact values of
$\theta $ and other critical exponents for self-avoiding branched
polymers in $D+2$ dimensions by relating them to the Yang-Lee
singularity of an Ising model in $D$ dimensions.  Various authors
\cite{D83,LF95,PF99} have also argued that the exponents of the
Yang-Lee singularity are related in simple ways to exponents for the
hard-core gas at the negative value of activity which is the closest
singularity to the origin in the pressure.  In this paper we consider
these models in the continuum and show that there is an exact relation
between the hard-core gas in $D$ dimensions and branched polymers in
$D+2$ dimensions.  We prove that the Mayer expansion for the pressure
of the hard-core gas is exactly equal to the generating function for
branched polymers.

Following \cite{F86}, we rewrite $c_{N}$ in a way that motivates the
continuum model we will study in this paper. Let $T$ be an abstract
tree graph on $N$ vertices labeled $1,\dots ,N$ and let $y =
(y_{1},\dots ,y_{N})$ be a sequence of distinct points in $\ZZ
^{D+2}$.  We say $y$ \emph{embeds} $T$ if $y_{ij}:= y_{i}-y_{j}$ has
length one for all \emph{edges} $\{i,j \}$ in the tree $T$. This
condition holds for $y$ if and only if it holds for any translate
$y'=(y_{1}+u,\dots ,y_{N}+u)$. Therefore it is a condition on the
class $[y]$ of sequences equivalent to $y$ under translation. Then
\[
c_{N} = \frac{1}{N!}\sum _{T,[y]} 1_{y \text{ embeds }T}
\]
\emph{Proof:} $\sum _{T} 1_{y \text{ embeds }T}$ is a symmetric
function of $y_{1},\dotsc ,y_{N}$ because a permutation $\pi$ of
$\{1,\dotsc ,N \}$ induces a permutation of tree graphs in the range
of the sum. Therefore, in the right-hand side of the claim, we can
drop the $\frac{1}{N!}$ and sum over representatives $(y_{1},\dotsc
,y_{N})$ of $[y]$ whose points are in lexicographic order.  Then the
vertices in the abstract tree $T$ may be replaced by points according
to $i \leftrightarrow y_{i}$ and the claim follows. \qed

The two systems to be related by dimensional reduction are

{\bf The Hard-Core Gas.}
Suppose we have ``particles'' at positions $ x_{1},\dots ,x_{N}$ in
a rectangle ${ \Lambda} \subset \RR ^{D}$.  Let $x_{ ij} = x_{i}-x_{j}$ and
define the  \emph{Hard-Core Constraint}:
\[
J ({ \{1,\dots ,N \}}, \mathbf{x}) = \begin{cases}
1 \text{ if all } |x_{ij}| \ge 1\\
0 \text{ otherwise }
\end{cases} .
\]
By definition, the \emph{ Partition Function} for the \emph{Hard-Core
Gas} is the following power series in ${\fz}$:
\begin{equation}\label{eq:def-zhc}
{ Z}_{\text{HC}} ({\fz}) =
\displaystyle{\sum _{N\ge 0}} \frac{z^{N}}{N!}
\int (d^Dx)^N \, J (\{1,\dots ,N \}, \mathbf{x}).
\end{equation}
where each $x_{i}$ is integrated over $\Lambda$. For $D=0$, $\Lambda$
is an abstract one-point space and the integrals can be omitted. Then,
the
hard-core constraint eliminates all terms with $N>1$ and the partition
function reduces to $1+{\fz}$.
\bigskip

{\bf Branched Polymers in the Continuum.}
A \emph{Branched Polymer} is a tree graph $T$ on vertices
$\{1,\dots ,N \}$ together with an { embedding} into $\RR ^{D+2}$,
{\emph{i.e.}} positions ${ y}_{i} \in \RR ^{D+2}$ for each $i=1,\dots ,N$, such
that
\begin{itemize}
\item[(1)] If $ij \in T$ then $|y_{ij}|=1$;
\item[(2)] If $ij \not \in T$ then $|y_{ij}|\ge 1$.
\end{itemize}
Define the weight $W (T)$ of a tree by
\begin{equation}\label{eq:def-wt}
W (T) := \int \prod _{ij \in T}
\begin{array}[t]{c}
\underbrace{d\Omega(y_{ij})}\\
\substack{\text{\tiny surface measure}\\
\text{\tiny on unit ball}}
\end{array}
\prod_{ij \notin T}
\one_{\{|y_{ij}|\geq 1\}}
\end{equation}
where the integral is over $\RR^{[D+2]N}/\RR^{D+2}$, or, more
concretely, $y_{1}=0$. If $N=1$, $W (T):=1$. The \emph{generating
function} for branched polymers is
\begin{equation}\label{eq:def-zbp}
{ Z_{\text{BP}}} ({\fz}) = \sum_{N=1}^{\infty} \frac{{\fz}^{N}}{N!}
\sum _{T \text{ \tiny on} \ \{1,\dots ,N \}} W (T).
\end{equation}

Our main theorem is
\begin{theorem} \label{thm:main} For all ${\fz}$ such that the right-hand
side converges absolutely,
\begin{equation*}
\lim_{\Lambda \nearrow \RR ^{D}} \frac{1}{|\Lambda |}\log
Z_{\rm{HC}} ({\fz})= -2 \pi Z_{\rm{BP}} \left(-\frac{{\fz}}{2\pi }\right) .
\end{equation*}
where $\lim$ is omitted when $D=0$.
\end{theorem}
The expansion of the left-hand side as a power series in $\fz$ is known
\cite{Rue69} to be convergent for $|\fz|$ small.

{\bf Consequences for Critical Exponents.} 
For $D=0,1$ the left-hand side can be computed exactly, and so we obtain exact formulas for the weights of polymers of size $N$ in dimension $d=D+2=2,3$:
\begin{corollary}\label{cor:exact}
\begin{equation}\label{identity}
\frac{1}{N!} \sum_{T {\rm \ on \ } \{1,\ldots,N\}} W(T) = 
\left\{\begin{array}{ll}
N^{-1}(2\pi)^{N-1} & {\rm \ if \ } d = 2 \\[2mm]
\frac{N^{N-1}}{N!}(2\pi)^{N-1} & {\rm \ if \ } d = 3
\end{array}\right. .
\end{equation}
\end{corollary}

\noindent \emph{Proof:} For $D=0$ the left-hand side of (\ref{identity}) is $\log(1+{\fz})$, and so
$$
Z_{\text{BP}}({\fz}) = - \, \frac{1}{2\pi} \, \log(1-2\pi {\fz}) = \sum^\infty_{N=1} \frac{1}{2\pi N}(2\pi {\fz})^N,
$$
which leads to the $d=2$ result.  For $D=1$, the pressure $\lim_{\Lambda \nearrow \RR ^{D}} |\Lambda |^{-1}\log
Z_{\text{HC}} ({\fz})$ of the hard-core gas is also computable (see \cite{HH63}, for example). 
It is the largest solution to $x e^{x} = {\fz}$ for ${\fz} > \tilde{{\fz}}_c := -e^{-1}$, and thus $2\pi Z_{\text{BP}}\left(- \, \frac{{\fz}}{2\pi}\right) = T(-{\fz})$.  Here $T({\fz})=-\text{Lambert}W(-{\fz})$ is the tree function, whose $N^{\text{th}}$ derivative at $0$ is $N^{N-1}$ (see \cite{CGHJK}).  Hence,
$$
Z_{\text{BP}}({\fz}) =  \frac{1}{2\pi}T(2\pi {\fz}) = \sum^\infty_{N=1} \frac{N^{N-1}}{2\pi N!}(2\pi {\fz})^N . 
$$
\qed

One can check directly from the definition above that the volume of the set of configurations available to dimers and trimers is indeed $\pi$, $4\pi^2/3$, respectively in $d=2$ and $2\pi, 6\pi^2$ respectively in $d=3$.  For larger values of $N$, Corollary \ref{cor:exact} describes a new set of geometric-combinatoric identities for disks in the plane and for balls in $\RR^3$.

From Corollary \ref{cor:exact} we see immediately that the critical activity ${\fz}_c$ for branched polymers in dimension $d=2$ is exactly $\frac{1}{2\pi}$, and that $\theta=1$.  For $d=3$, Stirling's formula may be used to generate large $N$ asymptotics:
$$
\frac{1}{N!} \sum_{T {\rm \ on \ } \{1,\ldots,N\}} W(T) = (2\pi)^{N-\frac{1}{2}}e^{-(N+1)}N^{-\frac{3}{2}}(1+O(N^{-1})).
$$
Hence ${\fz}_c=\frac{e}{2\pi}$ and $\theta=\frac{3}{2}$.

For $D=2$, the pressure of a gas of hard disks is not known, but if we assume the singularity at negative activity is in the same universality class as that of Baxter's model of hard hexagons on a lattice \cite{B82}, then the pressure has a leading singularity of the form $({\fz}-\tilde{{\fz}}_c)^{2-\alpha_{\text{HC}}}$ with $\alpha_{\text{HC}}={\frac{7}{6}}$ \cite{D83,BL87}.  We may define another critical exponent $\gamma_{\text{BP}}$ from the leading singularity of $Z_{\text{BP}}({\fz})$:
\begin{equation*}
\left( {\fz}\frac{d}{d{\fz}} \right) ^2Z_{\text{BP}}({\fz}) \sim ({\fz}-{\fz}_c)^{-\gamma_{\text{BP}}}, \text{or equivalently} \mbox{      } Z_{\text{BP}}({\fz}) \sim ({\fz}-{\fz}_c)^{2-\gamma_{\text{BP}}}.
\end{equation*}
Theorem \ref{thm:main} implies that the singularity of the pressure of the hard-core gas and the singularity of $Z_{\text{BP}}$ are the same, so 
\begin{equation}\label{eqn:gammaequalsalpha}
\gamma_{\text{BP}}=\alpha_{\text{HC}}. 
\end{equation}
Hence we expect that $\gamma_{\text{BP}}=\frac{7}{6}$ in dimension
$d=4$.  In general, if the exponent $\theta$ is well-defined, then it
equals $3-\gamma_{\text{BP}}$ by an Abelian theorem.  Thus $\theta$ should equal $\frac{11}{6}$ in $d=4$.

These values for $\theta(d)$ for $d=2,3,4$ agree with the Parisi-Sourlas relation 
\begin{equation}\label{eqn:ps}
\theta(d)=\sigma(d-2)+2 
\end{equation}
\cite{PS81} when known or conjectured values of the Yang-Lee edge exponent $\sigma(D)$ are assumed \cite{D83,C85} (see Section~\ref{section:background}).
Of course, the exponents are expected to be universal, so one should find the same values for other models of branched polymers (\emph{e.g.}, lattice trees) and also for animals.
\bigskip

{\bf A Generalization: Soft Polymers and the Yukawa Gas.}
We define
$$
Z_v({\fz}) = \sum_{N \geq 0} \ \frac{{\fz}^N}{N!} \int (d^Dx)^N \prod_{1 \leq i < j \leq N} e^{-v(|x_{ij}|^2)},
$$
where $x_i \in \Lambda \subset \mathbb{R}^D$ and $v(r^2)$ is a differentiable, rapidly decaying, spherically symmetric two-particle potential. The inverse temperature, $\beta$, has been included in $v$.  With $w(x) \equiv v(|x|^2)$, let us assume $\hat{w}(k) > 0$ for a repulsive interaction. Then there is a corresponding branched polymer model in $D+2$ dimensions with
\begin{equation}\label{equation:Wv}
W_v(T) := \int \prod_{ij\in T}
\left[- 2v'(|y_{ij}|^2)d^{D+2} y_{ij}\right]
\prod_{1 \leq i < j \leq N} e^{- v(|y_{ij}|^2)}.
\end{equation}
Note that by assumption, $v'(r^2)$ is rapidly decaying, so the monomers are stuck together along the branches of a tree. The polymers are softly self-avoiding, with the same weighting factor as for the Yukawa gas, albeit in two more dimensions. Defining, as before,
$$
Z_{\text{BP},v} = \sum_{N \geq 1} \frac{{\fz}^N}{N!} \sum_{T {\rm \ on \ } \{1,\ldots,N\}} W(T),
$$
we will prove

\begin{theorem}\label{theoremgeneralization}
For all ${\fz}$ such that the right-hand side converges absolutely,
\begin{equation}\label{eqn:softid}
\lim_{\Lambda \nearrow \mathbb{R}^D} \frac{1}{|\Lambda|} \log Z_v({\fz}) = -2\pi Z_{{\rm{BP}},v} \left( - \, \frac{{\fz}}{2\pi}\right).
\end{equation}
\end{theorem}

Note that by the sine-Gordon transformation \cite{KUH63,F76}
\begin{equation}\label{eqn:sg}
Z_v({\fz}) = \int \exp\left( \int dx \, \hat{{\fz}}e^{i\varphi (x)}\right) d \mu_{ w}(\varphi),
\end{equation}
where $d \mu_{ w}$ is the Gaussian measure with covariance $ w$, and $\hat{{\fz}} := {\fz} e^{ v(0)/2}$. Thus Theorem \ref{theoremgeneralization}
gives an identity relating certain branched polymer models and $-\hat{{\fz}}e^{i\varphi}$ field theories.
As discussed in Section~\ref{section:background}, an expansion of $-\hat{{\fz}}e^{i\varphi}$ about the critical point reveals an $i \varphi^3$ term (along with higher order terms), so we have a direct connection between branched polymers and the field theory of the Yang-Lee edge.
\bigskip

{\bf Green's Function Relations and Exponents.} Green's functions are
defined through functional derivatives as follows.  In the definition
(\ref{eq:def-zhc}) of the hard-core partition function $Z_{\text{HC}}$
each $dx_{j}$ is replaced by $dx_{j}\exp (h (x_{j}))$ where $h (x)$ is
a continuous function on $\Lambda$. Let $h = \alpha h_{1} + \beta
h_{2}$. Then there exists a measure $G_{\text{HC},\Lambda }
(dx_{1},dx_{2};\fz)$ on $\Lambda \times \Lambda$ such that
\[
\left.
\frac{\partial}{\partial \alpha } \frac{\partial}{\partial \beta }
\right |_{\alpha =\beta =0}
\log Z_{\text{HC}} = \int G_{\text{HC},\Lambda } (dx_{1},dx_{2};\fz)
h_{1} (x_{1})h_{2} (x_{2})
\]
This measure is called a density-density correlation or 2-point
Green's function because $G_{\text{HC},\Lambda
}(d\tilde{x}_{1},d\tilde{x}_{2};\fz)$ equals the correlation of $\rho
(d\tilde{x}_{1})$ with $\rho (d\tilde{x}_{2})$ where $\rho
(d\tilde{x}) = \sum \delta_{x_{j}}(d\tilde{x})$ is a random measure
interpreted as the empirical particle density at $\tilde{x}$ of the
random hard-core configuration $\{x_{1},\dotsc ,x_{N} \}$. 
(The underlying probability distribution on hard-core configurations is known as the Grand Canonical Ensemble;
$Z_{\text{HC}}(\fz)$ is its normalizing constant, \emph{c.f.} (\ref{eq:def-zhc}).)
For $\fz$
in the interior of the domain of convergence of the power series $Z_{\text{BP}}$, term
by term differentiation is legitimate and the weak limit as the volume
$\Lambda \nearrow \RR^{D}$ of $G_{\text{HC},\Lambda }
(dx_{1},dx_{2};\fz)$ exists. 
It is a translation-invariant measure which we write as $G_{\text{HC}}(dx;\fz)\,dx_{1}$, where $x=x_{2}-x_{1}$.  
These claims are easy consequences of our identities
but we omit the details since they are known \cite{Rue69}.

For branched polymers we define $\hat{W} (T)$ by changing the
definition (\ref{eq:def-wt}) of the weight $W (T)$ by (i) including an
extra Lebesgue integration over $y_{1}= (x_{1},z_{1}) \in \hat{\Lambda
}$, where $\hat{\Lambda}$ is a rectangle in $\RR^{D+2}$, and (ii)
inserting $\prod_{j} \exp (h (y_{j}))$ under the integral. Then
$\hat{Z}_{\text{BP}}$ is defined by replacing $W (T)$ by $\hat{W} (T)$
in (\ref{eq:def-zbp}).  We define the finite-volume branched polymer
Green's function as a measure by taking derivatives at zero with
respect to $\alpha$ and $\beta$ when $h = \alpha h_{1} + \beta
h_{2}$. The derivatives can be performed term by term and the infinite
volume limit as $\hat{\Lambda} \rightarrow \RR^{D+2}$ is easily
verified to be
\begin{equation} \label{eq:bp-green1}
G_{\text{BP}} (d\tilde{y}_{1},d\tilde{y}_{2};\fz) := 
\sum_{N=1}^{\infty} \frac{\fz^{N}}{N!}  \sum _{T \text{ \tiny on} \
\{1,\dots ,N \}} \int_{(\RR^{D+2})^N}\prod _{ij\in T} d\Omega(y_{ij})
\rho (d\tilde{y}_{1})\rho (d\tilde{y}_{2})
\end{equation}
where $\rho(d\tilde{y}) = \sum \delta_{y_{j}}(d\tilde{y})$. 
This can be written as $G_{\text{BP}}(d\tilde{y};{\fz}) d\tilde{y}_{1}$ where $\tilde{y}=\tilde{y}_{2}-\tilde{y}_{1}$.
%for $\tilde{y}\not =0$,
%\begin{equation} \label{eq:bp-green2}
%G_{\text{BP}}(d\tilde{y};{\fz}) = 
%\sum_{N=1}^{\infty} \frac{\fz^{N}}{(N-1)!}  \sum _{T \text{ \tiny on} \
%\{1,\dots ,N \}} \int_{\RR^{[D+2] (N-1)}}\prod _{ij\in T} d\Omega(y_{ij})
%\rho (d\tilde{y})
%\end{equation}
%The integral is over $x_{2},\dotsc ,x_{N}$ with $x_{1}=0$ because one
%$\rho $ was eliminated by integration, which also contributed a factor
%of $N$.

\begin{theorem}\label{thm:green}
If ${\fz}$ is in the interior of the domain of convergence of
$Z_{\text{BP}}$, then for all continuous compactly supported functions
$f$ of $x\in \RR^D$,
$$
\int_{\RR^{D}} f(x) G_{\rm{HC}}(dx;{\fz})=-2\pi\int_{\RR^{D+2}}f(x)
G_{\rm{BP}}\left(dy;-\frac{{\fz}}{2\pi} \right),
$$
where $y=(x,z)\in\RR^{D+2}$.
\end{theorem}

In effect, $G_{\text{HC}}$ can be obtained by integrating $G_{\text{BP}}$ over the two extra dimensions.  Note that
$G_{\text{BP}} (dy;\fz )$ is invariant under rotations of
$y$. Therefore, we can define a distribution $G_{\text{BP}}(t;{\fz})$ on
functions with compact support in $\RR^{+}$ by $\int f(t)
G_{\text{BP}}(t;{\fz}) dt = \int G_{\text{BP}} (dy;\fz
)f(|y|^2)$. $G_{\text{HC}}(t;{\fz})$ is defined analogously.  Then
Theorem~\ref{thm:green} implies that, in dimension $D\ge 1$,
\begin{equation}\label{eqn:GdG}
G_{\text{BP}}\left(t;-\frac{\fz}{2\pi}\right)=\frac{1}{2\pi^{2}}\frac{d}{dt} G_{\text{HC}}(t;\fz).
\end{equation}
where the derivative is a weak derivative. A similar theorem holds for
Green's functions associated with soft polymers and the Yukawa gas.

For $t>1$, which is twice the hard-core radius,
$G_{\text{HC}}(t;{\fz}) $ and $G_{\text{BP}}(t;{\fz}) $ are functions,
so one may define correlation exponents $\nu$ and $\eta$ from the
asymptotic form of the Green's functions as ${\fz} \nearrow {\fz}_c$.
The correlation length $\xi_{\text{HC}}({\fz})$ is defined from the
rate of decay of $G_{\text{HC}}$:
\begin{equation*}
\xi_{\text{HC}}({\fz})^{-1}:=\lim_{x\rightarrow \infty} -\frac{1}{x}
\log |G_{\text{HC}}(x^2;{\fz})|.
\end{equation*}
assuming the limit exists. Then the correlation length exponent
$\nu_{\text{HC}}$ is defined if $\xi_{\text{HC}}({\fz}) \sim
({\fz}-\tilde{{\fz}}_c)^{-\nu_{\text{HC}}}$ as ${\fz} \searrow
\tilde{{\fz}}_c := -2\pi {\fz}_c$.  One can then send $x\rightarrow
\infty$ and ${\fz} \searrow \tilde{{\fz}}_c$ while keeping $\hat{x} :=
x/\xi({\fz})$ fixed.  If there is a number $\eta_{\text{HC}}$ such
that the scaling function
\begin{equation}\label{eqn:defK}
K_{\text{HC}}(\hat{x}) := \lim_{x \rightarrow \infty, {\fz} \searrow
\tilde{{\fz}}_c} x^{D-2+\eta_{\text{HC}}}G_{\text{HC}}(x^2;{\fz})
\end{equation}
is defined and nonzero (at least for $\hat{x}>0$), then
$\eta_{\text{HC}}$ is called the anomalous dimension.  Similar
definitions can be applied in the case of branched polymers when
considering the behavior of $G_{\text{BP}}(y^2;{\fz})$ as ${\fz}
\nearrow {\fz}_c$ ($D$ is replaced with $d=D+2$ in (\ref{eqn:defK})).
Then (\ref{eqn:GdG}) implies that for $D \ge 1$,

\begin{eqnarray}
\xi_{\text{BP}}({\fz}) &=& \xi_{\text{HC}} \left( -\frac{{\fz}}{2\pi} \right)
\\
\nu_{\text{BP}} &=& \nu_{\text{HC}} \label{eqn:nu} \\[1mm]
\eta_{\text{BP}} &=& \eta_{\text{HC}} \label{eqn:eta} \\[1mm]
K_{\text{BP}}(\hat{x}) &=&\frac{1}{4\pi^{2}}
\left[
\hat{x}K'_{\text{HC}}(\hat{x})-(D-2+\eta_{\text{HC}})K_{\text{HC}}(\hat{x})
\right],
\end{eqnarray}
assuming the hard-core quantities are defined.

In conclusion, we see from (\ref{eqn:gammaequalsalpha}),
(\ref{eqn:nu}), (\ref{eqn:eta}) that the exponents
$\gamma_{\text{BP}}$, $\nu_{\text{BP}}$, $\eta_{\text{BP}}$ are equal
to their hard-core counterparts $\alpha_{\text{HC}}$,
$\nu_{\text{HC}}$, $\eta_{\text{HC}}$ in two fewer dimensions.  
If the relation $D\nu_{\text{HC}}=2-\alpha_{\text{HC}}$ holds for $D\le
6$ (hyperscaling conjecture) then a dimensionally reduced form of
hyperscaling will hold for branched polymers (\emph{c.f.} \cite{PS81}):
\begin{equation*}
(d-2)\nu_{\text{BP}} = 2-\gamma_{\text{BP}}.
\end{equation*}

For $D=1$ one has $\alpha_{\text{HC}}=\frac{3}{2}$,
$\eta_{\text{HC}}=-1$,
$K_{\text{HC}}(\hat{x})=-c\hat{x}^{-2}e^{-\hat{x}}$ \cite{BI01}.  Thus
our results prove that the branched polymer model
$Z_{\text{BP}}({\fz})$ has exponents $\gamma_{\text{BP}}=\frac{3}{2}$,
$\nu_{\text{BP}}=\frac{1}{2}$, $\eta_{\text{BP}}=-1$, and scaling
function
\begin{equation}\label{eqn:3dK}
K_{\text{BP}}(\hat{x})=c\hat{x}^{-1}e^{-\hat{x}}
\end{equation}
in three dimensions.  Equation (\ref{eqn:3dK}) was conjectured by
Miller \cite{Mil91}, under the assumption that a relation like
(\ref{eqn:GdG}) holds between branched polymers in $d=3$ and the
one-dimensional Ising model near the Yang-Lee edge (see section
\ref{section:background}).

For $D=2$, the conjectured value of $\alpha_{\text{HC}}$ is
$\frac{7}{6}$, as mentioned above.  Hyperscaling and Fisher's relation
$\alpha_{\text{HC}}=\nu_{\text{HC}}(2-\eta_{\text{HC}})$ then lead to
conjectures $\nu_{\text{HC}}=\frac{5}{12}$,
$\eta_{\text{HC}}=-\frac{4}{5}$.
Assuming these are correct, the results above imply the same values
for branched polymers in $d=4$.

In high dimensions ($d>8$) it has been proven that
$\gamma_{\text{BP}}=\frac{1}{2}$, $\nu_{\text{BP}}=\frac{1}{4}$,
$\eta_{\text{BP}}=0$ (at least for spread-out lattice models)
\cite{HS90},\cite{HS92},\cite{HvS00}.  While our results do not apply
to lattice models, they give a strong indication that the
corresponding hard-core exponents have the same (mean-field) values
for $D>6$.

\section{Background and Relation to Earlier Work}\label{section:background}
\setcounter{equation}{0}

In this section we consider theoretical physics issues raised by our results.  

Three classes of models are relevant to this discussion. {\em Branched polymers} and {\em repulsive gases} were defined in section \ref{section:introduction}. We also consider the {\em Yang-Lee edge} $h_\sigma(T)$, defined for the Ising model above the critical temperature as the first occurrence of Lee-Yang zeroes \cite{YL52} on the imaginary magnetic field axis. The density of zeroes is expected to exhibit a power-law singularity $g(h) \sim|h-h_\sigma(T)|^\sigma$ for $|\mbox{Im }h| > |\mbox{Im } h_\sigma(T)|$ \cite{KG71}.
This should lead to a branch cut in the magnetization, a singular part of the same form, and a free-energy singularity of the form $(h-h_\sigma(T))^{\sigma+1}$.  In zero and one dimension, the Ising model in a field is solvable and one obtains
$\sigma(0) = -1$, $\sigma(1) = - \, \frac{1}{2}$ \cite{F80}.
Above six dimensions, a mean-field model of this critical point should give the correct value of $\sigma$.  Take the
standard interaction potential
$$
V(\varphi) = \frac{1}{2} \, r \varphi^2 + u \varphi^4 + h \varphi,
$$
and let $h$ move down the imaginary axis. The point $\varphi_h$ where $V'(\varphi_h) = 0$ moves up from the origin, and when $h$ reaches the Yang-Lee edge $h_\sigma(r,u)$, one finds a critical point with $V'(\varphi_{h_c}) = V''(\varphi_{h_c}) = 0$.
One can easily see that $|\varphi_h-\varphi_{h_c}| \sim |h-h_c|^{1/2}$, which means that $\sigma=\frac{1}{2}$ in mean field theory.  Note that the expansion of $V(\varphi + \varphi_{h_c})$ then begins with a $\varphi^3$ term with purely imaginary coefficient. 

\bigskip
{\bf The repulsive-core singularity and the Yang-Lee edge.} The singularity in the pressure found for repulsive lattice and continuum gases at negative activity is known as the {\em repulsive-core singularity}.  Our Theorem \ref{thm:main} relates this singularity to the branched polymer critical point.  Poland \cite{P84} first proposed that the exponent characterizing the singularity should be universal, depending only on the dimension. Baram and Luban \cite{BL87} extended the class of models to include nonspherical particles and soft-core repulsions.  The connection with the Yang-Lee edge 
goes back to two articles: Cardy \cite{C82} related the Yang-Lee edge in $D$ dimensions to directed animals in $D+1$ dimensions, and Dhar \cite{D83} related directed animals in $D+1$ dimensions to hard-core lattice gases in $D$ dimensions. 
Another indirect link arises from the hard hexagon model which, as explained above, has a free-energy singularity of the form $({\fz}_c-{\fz})^{2-\alpha_{\text{HC}}}$ with $\alpha_{\text{HC}}=\frac{7}{6}$.  Equating $2-\alpha_{\text{HC}}$ with $\sigma+1$ leads to the value $\sigma(2)=-\frac{1}{6}$, which is consistent with the conformal field theory value for the Yang-Lee edge exponent $\sigma$ \cite{C85}.

More recently, Lai and Fisher \cite{LF95} and Park and Fisher \cite{PF99} assembled additional evidence for the proposition that the hard-core repulsive singularity is of the Yang-Lee class. In the latter article, a model with hard cores and additional attractive and repulsive terms was translated into field theory by means of a sine-Gordon transformation. When the repulsive terms dominate, a saddle point analysis leads to the $i\varphi^3$ field theory. We can simplify this picture by considering an interaction potential $w(x-y)$ with $\hat{w}(k) > 0$, $\int d^D k \, \hat{w}(k) < \infty$. Then the sine-Gordon transformation (\ref{eqn:sg}) leads to an interaction $-\hat{{\fz}}e^{i\varphi}$, where $\varphi$ is a Gaussian field with covariance $ w$ and $\hat{{\fz}} := {\fz} e^{ w(0)/2}$. In a mean-field analysis, $\varphi$ is assumed to be constant, and with $r = ( \hat{w}(0))^{-1}$ we obtain a potential
$$
V(\varphi) = -\hat{{\fz}}e^{i\varphi} + \frac{1}{2} \, r \varphi^2.
$$
If we put $\varphi = i x$, the saddle-point equation is
$$
\frac{\hat{{\fz}}}{r} = x e^x,
$$
which has two solutions for $-e^{-1} < \frac{\hat{{\fz}}}{r} < 0$. When $\hat{{\fz}} = \hat{{\fz}}_c = - \frac{r}{e}$, the two critical points coincide at $\varphi_{\hat{{\fz}}_c}$ such that $V'(\varphi_{\hat{{\fz}}_c}) = V''(\varphi_{\hat{{\fz}}_c}) =0$. Expanding about this point gives an $i\varphi^3$ field theory, plus higher-order terms. Complex interactions play an essential role here, since for real models, stability considerations prevent one from finding a critical theory by causing two critical points to coincide---normally at least three are needed, as for $\varphi^4$ theory.  Observe that for $\hat{{\fz}}-\hat{{\fz}}_c$ small and positive, the critical point $\varphi_{\hat{{\fz}}}$ satisfies $\varphi_{\hat{{\fz}}} - \varphi_{\hat{{\fz}}_{c}} \sim (\hat{{\fz}}-\hat{{\fz}}_c)^{\frac{1}{2}}$.  Hence this sine-Gordon form of the Yang-Lee edge theory also has $\sigma=\frac{1}{2}$ in mean field theory. 

\bigskip
{\bf Branched polymers and the Yang-Lee edge.} In \cite{PS81}, Parisi and Sourlas
connected branched polymers in $d$ dimensions with the
Yang-Lee edge in $d-2$ dimensions (see also \cite{F86}). Working with
the $n \rarrow 0$ limit of a $ \boldsymbol{\varphi}^3$ model, the
leading diagrams are the same as those of a $\varphi^3$ model in
imaginary random magnetic field. Dimensional reduction \cite{PS79}
relates this to the Yang-Lee edge interaction $i\varphi^3$ in two fewer
dimensions. The free-energy singularities should coincide, so
$2-\gamma_{\text{BP}}(d) = \sigma(d-2)+1$, therefore $\theta(d) = 3-\gamma_{\text{BP}}(d) =
\sigma(d-2)+2$. There are some potential flaws in this argument.  First, a
similar dimensional reduction argument for the Ising
model in a random (real) magnetic field leads to value of 3 for the lower critical dimension
\cite{PS79,KW81}, in contradiction with the proof of long-range order in $d=3$
\cite{I84, I85}. See \cite{BD98} for a discussion of this issue.
Second, nonsupersymmetric terms were discarded in the Parisi-Sourlas approach; also in Shapir's work on the lattice \cite{S83}.  Though irrelevant in the renormalization group sense, such terms could interfere with dimensional reduction.
Finding a more rigorous basis for dimensional reduction continues to be an important issue; for example Cardy's recent results on two-dimensional self-avoiding loops and vesicles \cite{C01} depend on a reduction of branched polymers to the zero-dimensional $i\varphi^3$ theory.

Our Theorems \ref{thm:main} and \ref{theoremgeneralization} provide an exact relationship between branched polymers and the repulsive-core singularity in two fewer dimensions.  When combined with the solid connection between
repulsive gases and the Yang-Lee edge, they leave little room to doubt the
Parisi-Sourlas claims for branched polymers.  In terms of exponents, we have $2-\gamma_{\text{BP}}(d)=2-\alpha_{\text{HC}}(d-2)=\sigma(d-2)+1$, and the Parisi-Sourlas relation (\ref{eqn:ps}) follows as above.
\bigskip

%\section{Main Results}\label{section:mainResults}

%\setcounter{equation}{0}

\section{A Fundamental Theorem of Calculus}\label{section:fund-th-calc}
\setcounter{equation}{0}

Suppose $ f (\mathbf{t})$ is a smooth function of compact support of
a collection $\mathbf{t} = (t_{ij}), (t_{i})$ of variables
\[\begin{array}[t]{c}
\underbrace{({ t_{ij}})_{1\le i<j\le N}}\\
\text{{\tiny { bond} variables}}
\end{array}
\text{ and }
\begin{array}[t]{c}
\underbrace{({ t_{i}})_{1\le i \le N}}\\
\text{{\tiny { vertex} variables}}
\end{array} .
\]
A subset $F$ of bonds $\{ij|1 \le i<j \le N\}$ is called a
\emph{Graph} on \emph{vertices} $\{1,\dots ,N \}$. A subset $R$ of
vertices is called a set of \emph{roots}. \emph{Forests} are graphs
that have no loops. Note that the empty graph is a forest by this
definition. The connected components of a forest are \emph{trees},
provided we declare that the graph with no bonds and just one vertex
is also a tree.

$f^{(F,R)} (\mathbf{t})$ denotes the derivative with respect to
the variables $t_{ij}$ with $ij \in F$ and $t_{i}$ with $i \in R$.
Let $ z_{1},\dots ,z_{N}$ be complex numbers, $z_{ ij} = z_{i}-z_{j}$
and set
\begin{equation*}
t_{ij} = |z_{ij}|^{2}, \hspace{5mm} t_{i} = |z_{i}|^{2} .
\end{equation*}

\begin{theorem}\label{thm:tree1} (Forest-Root Formula)
\begin{equation}\label{equation:frf}
f (\mathbf{0}) = \sum _{(F,R)} \int_{\CC ^{N}} f^{(F,R)}
(\mathbf{t})\left(\frac{d^{2}z}{-\pi } \right)^{N} ,
\end{equation}
\\
where $F,R$ is summed over all forests $F$ and all sets $R$ of roots
constrained by the condition that each tree in $F$ contains exactly one
root from $R$. $d^{2}z =du\,dv$ where $z = u+iv$.
\end{theorem}

This result is a generalization of Theorem 3.1 in
\cite{BW88}.
That paper and this one rely on
Lemma~\ref{lemma:supersymmetrylocalization}, an idea which is common  
to all of the papers \cite{PS79,PS80,L83,AB84}. The proof will be
given in Section~\ref{section:proofFR}.  The assumption of compact
support simplifies our discussion here. But having proven the theorem
in this case it holds, by taking limits, for any function which decays
to zero at infinity and whose first derivatives are continuous and
integrable.

\begin{figure}[h]
\centering
\includegraphics[width=.9\textwidth]{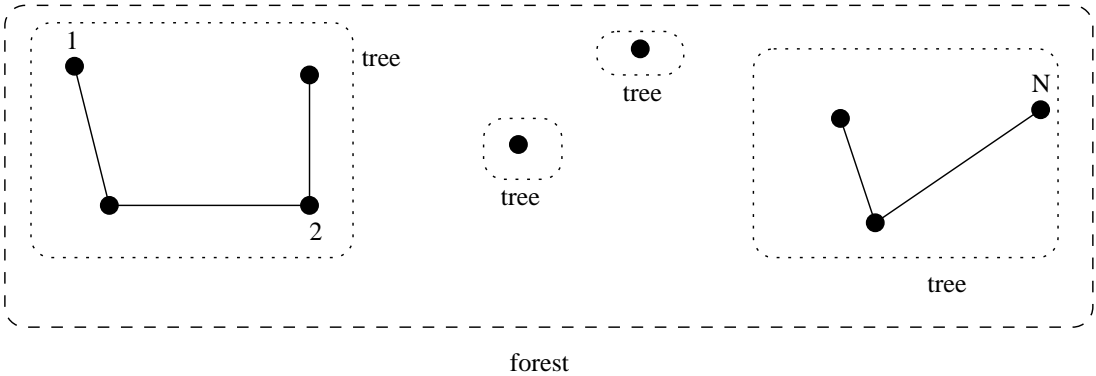}
\caption{Example of a forest}
\end{figure}

\section{A Tree Formula for Connected Parts}

\setcounter{equation}{0}

Let $J$ be a function on finite subsets $X$ of $\{1,2,\ldots\}$. The {\em connected part} of $J$ is a new function $J_c$ on finite subsets, uniquely defined by solving
\begin{equation}\label{eqn:connected}
J(X) = \sum_{\{X_{1},\ldots X_{n}\}, \ {\rm a \ partition \ of } \ X}
J_c(X_1)\cdots J_c(X_n)
\end{equation}
recursively, starting with $J_c(X) = J(X)$ if $|X| = 1$.

\begin{corollary}\label{corr:Jc}
Let $J(X) = J(X,\mathbf{t})$ depend on auxiliary parameters $t_{ij} \geq 0$ for each unordered pair $\{i,j\} \subset X$, $i \neq j$. Assume that
\begin{equation}\label{eqn:factor}
J(X \cup Y) = J(X)J(Y)
\end{equation}
for disjoint sets $X, Y$ whenever $t_{ij}$ is sufficiently large for all $i \in X$, $j \in Y$, or vice versa. Then
\begin{equation}\label{eqn:Jc}
J_c(X,\mathbf{0}) = \sum_{T {\rm \ on \ } X} \left(- \, \frac{1}{\pi}\right)^{N-1} \int_{\mathbb{C}^N/\mathbb{C}} J^{(T)}(X,\mathbf{t}),
\end{equation}
where $N = |X|$ denotes the number of vertices in $X$. The integral is over $z_i \in \mathbb{C}$, $i = 1,\ldots,N$ with simultaneous translations $z_i \rarrow z_i+c$ of all vertices factored out, and $t_{ij} = |z_i-z_j|^2$.
\end{corollary}

{\bf Remark.} This result was first proven for two-body interactions in \cite{BW88}. A simpler
proof based on the Forest-Root formula will be given here for arbitrary $J(X)$.  \bigskip

\noindent \emph{Proof:} Replace the labels $\{1,\dots ,N \}$ in the
Forest-Root formula by the elements of $X$.  Let $g$ be a smooth,
decreasing, compactly supported function with $g (0) =1$. Apply the
Forest-Root Formula (\ref{equation:frf}) to
\[
  f \big( (t_{ij}), (t_i ) \big) =
  J(X, (t_{ij} )) \prod _{i} g (\varepsilon t_i ),
\]
and let $\varepsilon >0$ tend to zero.  Then Corollary \ref{corr:Jc} is proved
by the following considerations:
\begin{figure}[h]
\centering
\includegraphics[width=.9\textwidth]{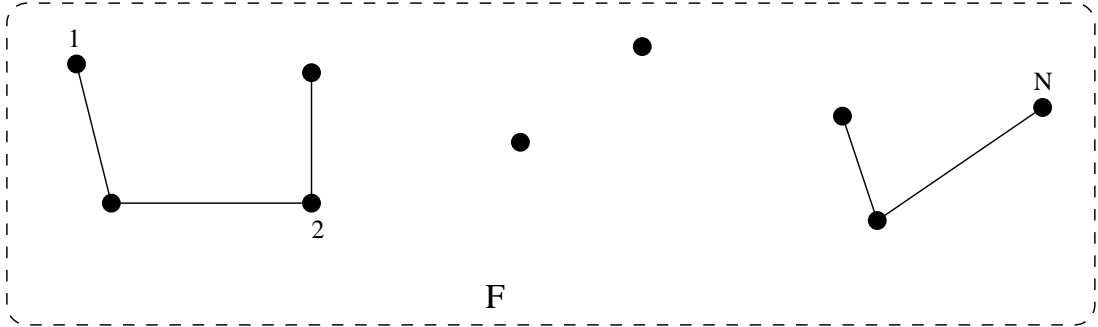}
  \caption{Partition on $X$ defined by a forest $F$}
\end{figure} 
\begin{enumerate}
\item A forest $F$ on a set of vertices $X$ uniquely determines a
partition of $X$, each subset being the vertices in one of the trees of $F$. Therefore,
\[
\sum _{F} (\cdots ) = \sum_{\{X_{1},\ldots ,X_{n}\}, \ {\rm a \
partition \ of } \ X} \hspace{2mm} \sum _{F, \ {\rm compatible \ with} \
\{X_{1},\ldots ,X_{n}\}  } (\cdots ).
\]

\item Consider any tree $T$ of $F$, and let $r$ be its root. There is
a factor $\varepsilon g'(\varepsilon t_r)$ from the root derivative at
$r$. Each of the other factors $g(\varepsilon t_i)$ for $i \neq r$ can be
replaced by $g(\varepsilon t_r)$ because hypothesis (\ref{eqn:factor})
makes any $t_{ij}$-derivative vanish for $t_{ij} \ge \text{const}$. 
This forces all $z_i$, with $i$ a vertex in $T$, to be equal
to within $O(1)$, and all $g(\varepsilon t_i)$ to be equal to within
$O (\varepsilon)$.

\item From the last item, and the sum over $r$ in $T$ (which comes
from the sum over $R$), there arises a factor $(-\varepsilon
N(T)/\pi)g'(\varepsilon t_r)g^{N(T)-1}(\varepsilon t_r)$ for each
tree.  (Here $N(T)=|T|+1$ denotes the number of vertices in $T$.)
This is a very ``flat'' probability density on
$\mathbb{C}$. The trees are distributed in $z$-space according to the
product of these probability densities.

\item As $\varepsilon \rightarrow 0$ the probability that any pair of
trees are within distance $o (\varepsilon ^{-1})$ tends to zero.
Thus, except for a set of vanishingly small measure, $J (X,t)$ factors 
into a product of terms, one for each tree on an $X_{i}$.

\item In the limit $ \varepsilon \to 0$, $\sum _{R}\int_{\CC ^{N}}
f^{(F,R)} (d^{2}z/ (-\pi ))^{X}$ equals the product over trees $T
\subset F$ of factors
\[
I (T) := \Big( - \frac{1}{\pi} \Big)^{ |T|}
\int_{\CC ^{N (T)}{ /\CC}} J^{(T)} (X_T,\mathbf{t}) ,
\]
where $X_T$ is the set of vertices in $T$.

\item
The sum over forests factors into independent sums over trees on each of the $X_i$.  
It follows that $\Sigma_{T {\rm \ on \ }X}I(T)$ solves the recursion (\ref{eqn:connected}); therefore it must be $J_c(X,\mathbf{0})$.
\end{enumerate}
\qed

\section{Proof of the Main Results}
\setcounter{equation}{0}

We prove Theorem \ref{thm:main} (relation between the hard-core gas
and branched polymers) by applying the tree formula for the connected
parts to the Mayer expansion:

\begin{theorem} \label{thm:mayer} \cite{M40} The formal power series for the logarithm
of the partition function is given by
\begin{equation} \label{eqn:me}
\log Z_{\rm{HC}}({\fz}) = \sum_{N \geq 1} \frac{{\fz}^N}{N!} \int (d^Dx)^N J_c(\{1,\ldots,N\},\mathbf{x}).
\end{equation}
\end{theorem}

\emph{Proof of Theorem~\ref{thm:main}:} The Hard-Core Constraint for
particles with labels in $X$ can be written as
$$
J(X,\mathbf{x}) = \prod_{ij \in X} \one_{\{|x_{ij}|^2\geq 1\}}.
$$
Let
$$
J(X,\mathbf{x},\mathbf{t}) = \prod_{ij \in X}
\one_{\{|x_{ij}|^2+t_{ij}\geq 1\}}.
$$
Replace each indicator function by a smooth approximation and apply
Corollary \ref{corr:Jc}, noting that
$$
\one_{\{|x_{ij}|^2+z_{ij}\bar{z}_{ij}\geq 1\}}
$$
is a hard-core condition in $D+2$ dimensions, and each
$t_{ij}$-derivative becomes $\frac{1}{2}$ surface measure when the
smoothing approximation is removed by taking a limit outside the
integrals. If we put $y_i = (x_i,z_i)$, a $(D+2)$-dimensional vector,
then, by Theorem~\ref{thm:mayer},
\begin{eqnarray} \label{eq:z-formula}
\lefteqn{
\log \, Z_{\text{HC}}({\fz})
} 
\nonumber \\[4mm]
& = & \sum_{N \geq 1} \
\frac{{\fz}^N}{N!} \sum_{T {\rm \ on \ } \{1,\ldots,N\}}
\left(- \, \frac{1}{\pi}\right)^{N-1} \int 
dx_{1}
\prod_{ij \in T}
\left[\frac{1}{2} \, d\Omega(y_{ij})\right]
\prod_{ij \notin T} \one_{\{|y_{ij}|\geq 1\}},
\end{eqnarray}
where the integral is over $(x_{1},\dotsc ,x_{N}) \in \Lambda^{N}$ and
$(z_{2},\dotsc ,z_{N})\in \RR^{2[N-1]}$ and $z_{1}=0$.  Consider the
integrations over $y_{2},\dotsc ,y_{N}$.  (i) By the monotone
convergence theorem the infinite volume limit as $\Lambda \rightarrow
\RR^{D}$ exists for each term in the sum over $N$. (ii) By translation
invariance the limit is independent of $x_{1}$ which is set equal to
zero. (iii) Division by $|\Lambda |$ cancels the remaining $dx_{1}$
integration over $\Lambda$. (iv) By absolute convergence of the sum
over $N$ the infinite volume limit can also be exchanged with the sum
over $N$. Theorem~\ref{thm:main} is proved. \qed

A similar argument can be used to prove Theorem \ref{theoremgeneralization}. 
It is necessary to relax the condition (\ref{eqn:factor}) in Corollary \ref{corr:Jc}.  The proof of Corollary \ref{corr:Jc} remains valid if $J$ has a clustering property:
$$
J(X \cup Y)\rightarrow J(X)J(Y) {\mbox{  when all }} t_{ij}\rightarrow\infty {\mbox{ for }} i\in X, j\in Y,
$$
and if a similar statement holds with $(\frac{\partial}{\partial t})^F$applied.
This is satisfied for
$$
J(X,\mathbf{x},\mathbf{t}) = \prod_{ij \in X} e^{- v(|x_{ij}|^{2}+t_{ij})} =
\prod_{ij\in X} e^{- v(|y_{ij}|^2)},
$$
provided $v$, $v'$ vanish at infinity.  We further assume that $v'(|y|^2)$ is an integrable function of $y\in \RR ^{D+2}$.
When evaluating $J^{(T)}$ in (\ref{eqn:Jc}), the factors $- v'(|y_{ij}|^2)$ ensure convergence of (\ref{equation:Wv}).  An extra factor of $2$ has been inserted in (\ref{equation:Wv}) so that the combination $-\frac{{\fz}}{2\pi}$ appears in (\ref{eqn:softid}).
\qed

\emph{Proof of Theorem \ref{thm:green}:} If $\prod dx_{j}$ is replaced
by $\prod \exp (h (x_{j}))dx_{j}$ in the definition (\ref{eq:def-zhc})
of $Z_{\text{HC}}$ the proof of (\ref{eq:z-formula}) generalizes to
\begin{eqnarray*} 
\lefteqn{
\log \, Z_{\text{HC}}(\fz e^{h}) 
}
\\[4mm]
& = &
\sum_{N \geq 1} \
\frac{{\fz}^N}{N!} \sum_{T {\rm \ on \ } \{1,\ldots,N\}}
\left(- \, \frac{1}{\pi}\right)^{N-1} \int 
dx_{1}
\prod_{ij \in T}
\left[\frac{1}{2} \, d\Omega(y_{ij})\right]
\prod_{ij \notin T} \one_{\{|y_{ij}|\geq 1\}}\prod_j \exp (h (x_{j})),
\end{eqnarray*}
where the integral is over $(x_{1},\dotsc ,x_{N}) \in \Lambda^{N}$ and
$(z_{1},\dotsc ,z_{N})\in \RR^{2N}/\RR^{2}$.  We differentiate with
respect to $\alpha ,\beta$ at zero with $h = \alpha h_{1}+\beta h_{2}$
and $h_{i}$ compactly supported. The left-hand side becomes the 
finite-volume Green's function $G_{\text{HC},\Lambda }
(d\tilde{y}_{1},d\tilde{y}_{2};\fz)$ integrated against the test
functions $h_{1} (\tilde{x}_{1})$ and $h_{2} (\tilde{x}_{2})$, and the
right-hand side becomes 
\[
-2\pi \sum_{N=1}^{\infty} \frac{1}{N!} \left(-\frac{\fz}{2\pi}\right)^N \sum _{T \text{ \tiny on} \
\{1,\dots ,N \}} \int \prod _{ij\in T} d\Omega(y_{ij})
\rho (h_{1})\rho (h_{2}) ,
\]
where $\rho(h) = \sum h (x_{j})$ and the integral is over $x_{j} \in
\Lambda$ for $j=1,\dotsc ,N$ and $(z_{1},\dotsc ,z_{N})\in
\RR^{2N}/\RR^{2}$. The integration over $\RR^{2N}/\RR^{2}$ can be rewritten
as an integral over $\RR^{2N}$ by replacing $\rho (h_{1})$ by $\sum h_1
(x_{j})\delta (z_{j})$.   The test functions localize at least one $x_{j}$ in
their support, so as in the proof of Theorem~\ref{thm:main}
the infinite volume limit $\Lambda
\rightarrow \RR^{D}$ exists by the monotone and dominated convergence
theorems. 
The limit is easily verified to be
\[
\int G_{\text{BP}} (d\tilde{y}_{1},d\tilde{y}_{2};\fz) 
h_{1} (\tilde{y}_{1}) \delta (\tilde{z}_{1})
h_{2}(\tilde{y}_{2})
\]
and this proves Theorem~\ref{thm:green}. \qed

The generalization of Theorem \ref{thm:green} to $n$-point functions is straightforward.  On the branched polymer side of the identity all but one of the points are integrated over two extra dimensions.  In Fourier space, this means that branched polymer Green's functions equate to hard-core Green's functions when all components of momenta for the extra dimensions are set to zero.

\section{Proof of the Forest-Root Formula}\label{section:proofFR}
\setcounter{equation}{0}

Define the differential forms
\begin{gather*}
{ \tau }_{ij} = z_{ij}\zbar _{ij} + dz_{ij}d\zbar _{ij}/ (2\pi i) ,\\
{ \tau }_{i} = z_{i}\zbar _{i} + dz_{i}d\zbar _{i}/ (2\pi i) .
\end{gather*}
Forms are multiplied by the wedge product. Suppose $g (t_{1})$ is a
smooth function on the real line. Then we define a new form by the
Taylor series
\[
{ g} (\tau _{1}) = g (z_{1}\zbar _{1}) + g' (z _{1}\zbar _{1})dz
_{1}d\zbar _{1}/ (2\pi i) ,
\]
which terminates after one term because all higher powers of $dz _{1}
d\zbar _{1}$ vanish. More generally, given any smooth function of the
variables $(t_{i}), (t_{ij})$, we define $g (\tau )$ by the analogous
multivariable Taylor expansion.  By \emph{definition}, integration
over $\CC ^{N}$ of forms is zero on all forms of degree not equal to
$2N$.

The following lemma exploits the supersymmetry of this setup to localize the evaluation of integrals on $\CC ^{N}$ to the origin.  It will be proven in section
\ref{section:equivariant}.
\begin{lemma}\label{lemma:supersymmetrylocalization}(supersymmetry and localization)
For $f$ smooth and compactly supported, 
\begin{equation*}
\int_{\CC ^{N}} f (\tau ) = f(0). 
\end{equation*}
\end{lemma}

Let $G$ be any graph on vertices $\{1,2,\dots ,N \}$.  Define
\[
(dzd\zbar )^{G} = \prod _{ij \in G}dz_{ij}d\zbar _{ij} ,
\]
and analogously, for $R$ any subset of vertices
\[
(dzd\zbar )^{R} = \prod _{i \in R}dz_{i}d\zbar _{i} .
\]
The Taylor series that defines $f (\tau )$ can be written in the form
\[
f (\tau ) = \sum _{G,R} f^{(G,R)} (z\zbar ) \left(\frac{dzd\zbar}{2\pi i}\right)^{G}
\left(\frac{dzd\zbar}{2\pi i}\right)^{R},
\]
where $G$ is summed over all graphs and $R$ is summed over all subsets
of vertices.
\begin{itemize}
\item $(dzd\zbar )^{G} =0 $ if $G$ contains a loop $ L$ because
$\sum _{ij\in L}z_{ij}=0$. Therefore $G$ must be a forest.
\item $(dzd\zbar )^{F} (dzd\zbar )^{ R}$ has degree $2N$ iff $R$ has
the same number of vertices as there are trees in $F$. This is because
a tree on $m$ vertices has $m-1$ lines.
\item Each tree contains exactly one vertex from $R$, because, if $T$
is a tree which include two vertices $a,b$ from $R$ then $(dzd\zbar )^{T}
(dzd\zbar )^{ R}=0$ since $z_{a} - z_{b}$ is a sum of $z_{ij}$ over
$ij$ in the path in $T$ joining $a$ to $b$.
\end{itemize}
By these considerations Theorem \ref{thm:tree1} is reduced to

\begin{lemma}\label{lemma:no-jacobian}
$$
\left(\frac{dzd\zbar}{2\pi i}\right)^{F}
\left(\frac{dzd\zbar}{2\pi i}\right)^{R} = \frac{d^{2}z_{1}}{-\pi } \dots
\frac{d^{2}z_{N}}{-\pi } .
$$
\end{lemma}

\begin{figure}[h]
\centering
\includegraphics[width=.75\textwidth]{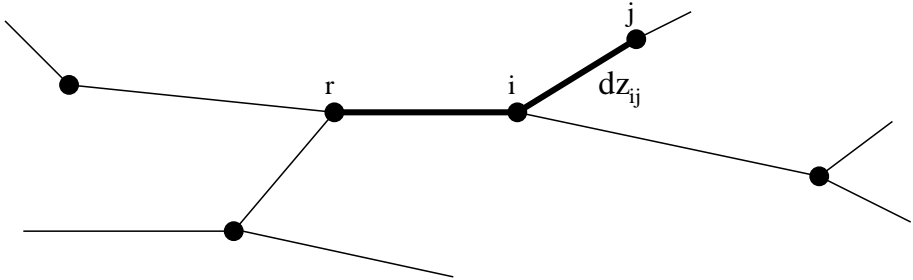}
\caption{Unique path property}
\end{figure} 

\noindent \emph{Proof:} Suppose, by changing the labels if necessary,
that vertices are labeled in such a way that as one traverses any path
in $F$ starting at a root $r$, the vertices one encounters have
increasing labels. Thus, in the figure, $r<i<j$.  Let $j=N$. Then $z
_{ij}$ may be replaced by $z _{j}$ because
\[
dz _{j} = \sum _{kl {\text{\tiny{\mbox{ in path}}}}} dz _{kl} + dz
_{\text{\tiny root}} ,
\]
and $ (dzd\zbar )^{ F} (dzd\zbar )^{R}$ already contains
$dz_{\text{\tiny root}}$ and the other terms in the path. This argument
may be repeated for $j$ decreasing through $N-1,N-2,\dots 1$. The
lemma then follows from the fact that if $z = x+iy$, then $dz d\bar{z}
= -2i \, dx \, dy$.  \qed

\section{Equivariant Flows and Dimensional Reduction}\label{section:equivariant}
\setcounter{equation}{0}

{\em Proof of Lemma \ref{lemma:supersymmetrylocalization} (the supersymmetry/localization lemma).}
We prove the identity $\int_{\mathbb{C}^N} f(\tau) = f(0)$ first in the special case $N=1$ so that $f(\tau) = f(\tau_1)$:
\begin{gather*}
\int_{\CC } f (\tau _{1} ) =
\begin{array}[t]{c}
\underbrace{\int_{\CC } f (z _{1}\zbar _{1} )}\\
0 \text{ \tiny   by definition}
\end{array}
+ \int_{\CC } f
'(z _{1}\zbar _{1} ) dz _{1} d\zbar _{1} / (2\pi i)\\
= -2 \int _{0}^{\infty } f' (r^{2})rdr = f (0) .
\end{gather*}
Note that this proof for $N=1$ generalizes to the case where $f$ depends
only on vertex variables $\tau_{1},\dots ,\tau_{N}$.  The remaining argument
is a reduction to this case borrowing ideas from the proof of the
Duistermaat-Heckmann theorem in \cite{AB84}, as explained in \cite{Wit92}.

There is a flow on $\CC ^{N}$: $z_{j} \longmapsto e^{-2\pi i\theta
}z_{j}$.  Let $V$ be the associated vector field and let $i_{V}$ be
the associated interior product which is an anti-derivation on
forms. By definition $i_{V} dz_{i} = - 2 \pi i z $ and $i_{V}
d\zbar_{i} = 2 \pi i \zbar$. Let $\cL _{V}$ be the associated Lie
derivative on forms. It is a derivation on forms,
\[
\cL _{V} d z_{i} = \frac{d}{d\theta}\bigg|_{\theta =0} d (e^{-2\pi
i\theta }z _{i}) = - 2\pi i d z _{i}
\]
and $\cL _{V} d \zbar _{i} = 2\pi i d \zbar _{i}$.  Define the
antiderivation $Q = d + i_{V}$ and note that $Q^{2} = d i_{V} + i_{V}
d = \cL _{V}$ by Cartan's formula for $\cL _{V}$.
\begin{enumerate}
\item $\tau _{ij} = Qu_{ij}$ with $u_{ij} = z_{ij} d\zbar _{ij}/ (2\pi
i)$ .
\item $Q\tau _{ij} = Q^{2}u_{ij} = \cL _{V}u_{ij} =0$ because $u_{ij}$
is invariant under the flow.
\item For any smooth function $g$, $Qg (\tau ) = \sum g^{(ij)} (\tau )
\, Q\tau _{ij} = 0$ .
\item Fix any bond $ij$ and define $f ({\beta  },\tau )$ by
replacing $\tau _{ij}$ in $f (\tau )$ by ${ \beta} \tau _{ij}$. Then
\[
\frac{d}{d\beta } f (\beta ,\tau ) = d \mbox{ (some form)} +
\mbox{(a form of degree $< 2N$)} ,
\]
because the $\beta$-derivative is $f^{(ij)}(\beta,\tau  )\tau _{ij}$
which equals
\[
f^{(ij)} (\beta,\tau  ) Qu = Q (f^{(ij)} (\beta,\tau ) u) ,
\]
and $Q= d + i_{V}$ and $i_{V}$ lowers the degree by one.
\item $\frac{d}{d\beta } \int_{\CC ^{N}} f (\beta ,\tau )=0$ because
the integral annihilates the part of lower degree and also annihilates
the part in the image of $d$ by Stokes' theorem.
\end{enumerate}
Thus every $\tau _{ij}$ in $f$ can be deformed to $0$ and
Lemma \ref{lemma:supersymmetrylocalization} is reduced to the case where $f$ is a
function only of $(\tau _{1}, \tau _{2},\dots ,\tau _{N})$. \qed

\begin{center}
\large{\textbf{Acknowledgment}}
\end{center}
We thank Gordon Slade for helpful comments and questions that improved the paper.

%\bibliography{brpol}

\newcommand{\etalchar}[1]{$^{#1}$}

\end{document}